\documentclass[aip,jcp,reprint]{revtex4-1}

\usepackage{makecell}
\usepackage{array}
\usepackage{physics}
\usepackage{lmodern}
\usepackage{graphicx}
\usepackage{amsmath}
\usepackage{amssymb}
\usepackage{float}
\usepackage{color}
\usepackage{xspace}

\graphicspath{{paper0_figures/}}

\newcommand{\para}{\textit{para}-H$_2$\xspace}
\newcommand{\spara}{solid \textit{para}-H$_2$\xspace}
\newcommand{\fcc}{\textit{fcc}\xspace}
\newcommand{\hcp}{\textit{hcp}\xspace}
\newcommand{\ANG}{\text{\AA}\xspace}
\newcommand{\etal}{\textit{et al.}\xspace}
\newcommand{\apriori}{\textit{a priori}\xspace}

\newcommand{\vs}{\textit{vs}\xspace}

\usepackage{silence}
\begin{document}

\title{Equation of State and First Principles Prediction of the Vibrational Matrix Shift of Solid Parahydrogen}

\author{Alexander Ibrahim}
\affiliation{Department of Physics and Astronomy, University of Waterloo, 200 University Avenue West, Waterloo, Ontario N2L 3G1, Canada}
\affiliation{Department of Chemistry, University of Waterloo, 200 University Avenue West, Waterloo, Ontario N2L 3G1, Canada}

\author{Lecheng Wang}
\affiliation{Department of Chemistry, University of Waterloo, 200 University Avenue West, Waterloo, Ontario N2L 3G1, Canada}

\author{Tom Halverson}
\affiliation{Department of Chemistry, University of Waterloo, 200 University Avenue West, Waterloo, Ontario N2L 3G1, Canada}

\author{Robert J Le Roy}
\thanks{Deceased.}
\affiliation{Department of Chemistry, University of Waterloo, 200 University Avenue West, Waterloo, Ontario N2L 3G1, Canada}

\author{Pierre-Nicholas Roy}
\email{pnroy@uwaterloo.ca}
\affiliation{Department of Chemistry, University of Waterloo, 200 University Avenue West, Waterloo, Ontario N2L 3G1, Canada}

\begin{abstract}
We generate the equation of state (EOS) of solid parahydrogen (\para) using a path-integral Monte Carlo (PIMC) simulation based on a highly accurate first-principles adiabatic hindered rotor (AHR) potential energy curve for the \para dimer. The EOS curves for the \fcc and \hcp structures of \spara near the equilibrium density show that the \hcp structure is the more stable of the two, in agreement with experiment. To accurately reproduce the structural and energy properties of \spara, we eliminated by extrapolation the systematic errors associated with the choice of simulation parameters used in the PIMC calculation. We also investigate the temperature dependence of the EOS curves, and the invariance of the equilibrium density with temperature is satisfyingly reproduced. The pressure as a function of density, and the compressibility as a function of pressure, are both calculated using the obtained EOS and are compared with previous simulation results and experiments. We also report the first ever \apriori prediction of a vibrational matrix shift from first-principles two-body potential functions, and its result for the equilibrium state agrees well with experiment.
\end{abstract}

\maketitle

\section{Introduction} \label{sub:intro}

For the past several decades, the properties and applications of \spara have been studied extensively. Its weak intermolecular interactions and low mass make \para crystals very ``soft'' and give it many interesting quantum properties.\cite{ph2solex:66nosa, ph2solex:80silv} Solid \para is one of the simplest examples of a quantum crystal.\cite{ph2solex:60gush, ph2solex:66barr, ph2solex:67bost, ph2solex:70schne, ph2solth:91Zoppi, ph2solex:92momo, ph2solex:93oka, ph2solth:02Zoppi, ph2solth:06Oper, ph2solth:06Tom} Also, as a spinless boson, \para has proven itself to be a viable candidate for superfluidity, both spectroscopically and theoretically.\cite{suph2:00greb, suph2:01agreb, suph2:01bgreb, suph2:02greb, suph2:03greb, suph2:05peas, suph2:10greb, suph2:10hli, suph2:13atoby, suph2:13btoby} For example, computer simulations have shown that ``glassy'' \para posesses superfluid properties.\cite{ph2solth:12Osy} Experimentally, \spara is an excellent accommodating host for impurity spectroscopy, owing to its relatively large lattice constant and the spherical symmetry of its \textit{J}~$ = 0$ rotational level.\cite{doh2solex:93weli, doh2solex:98zhang, doh2solex:04momo, doh2solex:13faja}

A large amount of research has been dedicated to characterizing the thermodynamics of \para. For example, relative to its value in the gas phase, the frequency shift of the Raman Q$_1$(0) transition at $ T = 4.2 $~K was measured by Oka \etal to be $ -11.43 $~cm$^{-1}$ in \spara.\cite{ph2solex:93oka} This is more than twice the vibrational frequency shift observed in large \para clusters.\cite{ph2solex:93oka,ph2clu:04teje,ph2clu:14Nabi} To date, no theoretical studies have been able to address this discrepancy. Another quantity attracting wide interest is the isotherm curve of \para, particularly at $ T = 4.2 $~K.\cite{ph2solex:63good, ph2solex:75dura, ph2solth:78silv, ph2solex:79drie} Silvera's review covers the subject far beyond the scope of the discussion here.\cite{ph2solex:80silv} The zero pressure molar volume and high compressibility of \spara at $ 4.2 $~K both make a strong case that it is a quantum solid at that temperature.\cite{ph2solex:80silv} Further support for this idea comes from Schuch \etal, who have shown that \spara has a hexagonal close-packing (\hcp) structure in its equilibrium state.\cite{ph2solex:68schu}

Several computational studies have attempted to provide insights to the experiments outlined above. In $ 1996 $, Cheng and Whaley performed Diffusion Monte Carlo (DMC) calculations on \spara at equilibrium density\cite{ph2solth:96chen} using both the Buck and Silvera-Goldman (SG) empirical two-body pair potentials.\cite{h2pes:78sg, h2pes:83buck} The SG term includes a repulsive three-body correction term, while the Buck potential does not. In $ 2006 $, Operetto and Pederiva performed a similar DMC simulation over a wider density range. They predicted the equation of state (EOS) of \spara and confirmed that its preferred crystal structure is \hcp, matching the experimental consensus.\cite{ph2solth:06Oper} Using path-integral Monte Carlo (PIMC) calculations, Osychenko \etal investigated the onset of superfluidity of glassy \para.\cite{ph2solth:12Osy} Their research highlighted the sensitivity of computational results to the choice of potential energy surface (PES), and emphasized the need for more accurate PES's for \para that better account for subtle quantum effects. For example, understanding the influence of finite size effects is paramount to accurately reproducing high-resolution spectroscopic measurements.

Recently, by solving a rovibrational Hamiltonian for \para monomers based on the $6$D Hinde H$_2$--H$_2$ dimer PES,\cite{h2pes:08hinde} Schmidt \etal reported first-principles adiabatic hindered rotor (AHR) reduced-dimension potentials for \para--\para and \textit{ortho}-D$_2$--\textit{ortho}-D$_2$, for both the internal vibrational ground and excited states.\cite{ph2clu:15schm} These potentials were able to reproduce the experimentally observed infrared vibrational shifts for clusters of various sizes.\cite{ph2clu:14Nabi} The work presented in this paper uses these \para--\para PES's, which we refer to as the Faruk-Schmidt-Hinde (FSH) potential. We should note that, because information about the ro-vibrational states of the \para monomers has been averaged out of the aforementioned empirical Buck and SG potentials, they cannot predict vibrational band shifts.

Also presented in this paper are results from a PIMC program created to simulate a \spara crystal using both face-centred cubic (\fcc) and \hcp structures.\cite{pi:14tobya} For the EOS at $ T = 4.2 $~K, to help eliminate the systematic errors inherent to PIMC simulations, the number of time slices in the simulation and the number of \para molecules inside the periodic boundary cell were extrapolated to infinity. Using this EOS we computed the pressure as a function of density and the compressibility as a function of pressure. We also computed the infrared vibrational frequency shift as a function of density. This was made possible due to the aforementioned work of Schmidt \etal, whose PES includes intramolecular stretching. To investigate the temperature dependence of the EOS and vibrational matrix shift, we also perform several simulations using a finite number of time slices and particles at several different isotherms.

The remainder of this paper is organized as follows: In Sec.~\ref{sec:theory} we describe our theoretical formalism of the \spara system and the PIMC methodology. In Sec.~\ref{sec:numerics} we provide an overview of our numerical approach to the elimination of systematic errors in our PIMC simulations. We present and discuss the results in Sec.~\ref{sec:results}, and end with concluding remarks in Sec.~\ref{sec:conclusion}.

\section{Theory\label{sec:theory}}
\subsection{Hamiltonian, pairwise approximation, and symmetry} \label{sub:hamiltonian}

A lot of effort has gone into studying the quantum nature of \spara. However, because the bulk effects of \spara only appear for large numbers of particles, exact calculations of these properties are intractable. We can make calculations more manageable by using a series of adiabatic approximations in a Born-Oppenheimer-like (BOL) manner, similar to common methods used in electronic structure. Coupled with a pairwise approximation, this simple methodology allows us to compute many thermodynamic properties while still using a first-principles PES.

The Hamiltonian for a system of $ N $ interacting \para molecules can be expressed as
\begin{equation} \label{eq:hamiltonian}
	\hat{H}(\vb{x}_1, \vb{p}_1, ..., \vb{x}_N, \vb{p}_N) = \frac{1}{2 \mu} \sum^{N}_{i = 1} \vb{p}_i^2 + U(\vb{x}_1, ..., \vb{x}_N) ,
\end{equation}
\noindent
where $ \mu $ is the mass of each \para molecule, and $ \vb{p}_i $ and $ \vb{x}_i $ are the momentum and position of the \textit{i}$^{\rm th}$ molecule, respectively. Here, we use a pairwise additive potential $ U $ as an approximation to the full, many body potential. Originally derived by Faruk \etal,\cite{ph2clu:14Nabi} this pairwise potential is computed by treating the \para--\para separation as the ``slow'' coordinate in the BOL framework. The resulting Hamiltonian is
\begin{equation} \label{eq:pairwise_hamiltonian}
	\hat{h}(R_{12}) = \hat{h}_1 + \hat{h}_2 + V^{6D}(r_1, r_2, \theta_1, \theta_2, \phi; R_{12}) ,
\end{equation}
\noindent
where $ V^{6D} $ is the \para--\para rotor interaction potential produced by Hinde, and $ \hat{h}_1 $ and $ \hat{h}_2 $ are the rovibrational Hamiltonians for the individual \para molecules, respectively.\cite{h2pes:08hinde} We then solve Eq.~(\ref{eq:pairwise_hamiltonian}) at a series of intermolecular spacings $ R_{12} = \left| \vb{x}_1 - \vb{x}_2 \right| $ for the ground state eigenvalues, which in turn constitute the given pairwise potential. To compute the eigenvalues we expand Eq.~(\ref{eq:pairwise_hamiltonian}) in a product basis of the molecular eigenfunctions. This yields a basis set of the form
\begin{equation} \label{eq:BOL_hamiltonian_basis}
	\ket{I} = \bigotimes^{2}_{i = 1} \ket{l_i m_i \nu_i} = \ket{l_1 m_1 \nu_1 l_2 m_2 \nu_2} ,
\end{equation}
\noindent
where $ I = \{ l_1, m_1, \nu_1, l_2, m_2, \nu_2 \} $ is the composite index.

In general, because the individual \para molecules are indistinguishable, the (\para)$_N$ system must exhibit exchange symmetry. We know that the Raman modes have $ A_1 $ character - i.e. we are only interested in the eigenstates from the totally symmetric irrep. However, because we are using a BOL approximation to separate the centre-of-mass motion from the internal rotation, we cannot explicitly apply the exchange symmetry to our basis. The full exchange operation permutes \textit{all} coordinate indices, including $ R_{12} \rightarrow R_{21} $, and because $ R_{12} $ is a parameter of Eq.~(\ref{eq:pairwise_hamiltonian}), it is unaffected by the permutation operator, i.e.~$ \hat{P}_{12} \ket{l_1 m_1 \nu_1 l_2 m_2 \nu_2} = \ket{l_2 m_2 \nu_2 l_1 m_1 \nu_1}$. This presents a problem because our criterion for the BOL separation is simply energetic ordering, and we cannot guarantee the character of the ground state. To ensure that all the chosen states in the final pairwise potential have the proper $ A_1 $ symmetry, we define a projection operator for the totally symmetric irrep,
\begin{equation} \label{eq:projection_operator}
	\Delta_{+} = \frac{1}{2} \left( \hat{I} + \hat{P}_{12} \right) ,
\end{equation}
\noindent
and apply it to Eq.~(\ref{eq:pairwise_hamiltonian}),
\begin{equation}
	\hat{h}_{+}(R_{12}) = \Delta_{+}^{T} \hat{h} \Delta_{+} .
\end{equation}
\noindent
At first glance, this procedure turns out to be equivalent to starting with a symmetrized basis. However, this is only the case for the pair, since the only parametric exchange term is $ R_{12} $, which is equivalent to $ R_{21} $. In general, this does not hold for an arbitrary $ R_{ij} $, but such an issue is beyond the scope of this paper.

Another quantity of interest is the ``vibrational matrix shift'', the change in the ground state energy of the system after the introduction of an internal quantum of vibration. Experiments show that the energy spacing of this vibrational excitation is very large compared to the internal rotational energy level spacing,\cite{rovib:59kran, rovib:68kran, adiab:09krz} and thus the problem can be treated adiabatically. By taking the total vibrational quantum to be the sum of the individual quanta, $ \nu_t = \nu_1 + \nu_2 $, we can create two surfaces by restricting $ \nu_t = 0 $ or $ \nu_t = 1 $. We then compute the ground state energy of Eq.~(\ref{eq:pairwise_hamiltonian}) as a function of the pairwise distance, under the two vibrational restrictions,
\begin{equation} \label{eq:restrictions}
	\mel{I}{\hat{h}(R_{ij})}{J} \rightarrow E_0^{\nu_t}(R_{ij}) \rightarrow V^{\nu_t}(R_{ij}) .
\end{equation}
\noindent
We define $ V^1(r) $ to be the potential energy function of a \para dimer where one molecule is in the $ \nu = 1 $ state and the other is in the $ \nu = 0 $ state, and $ V^0(r) $ to be its counterpart where both \para molecule are in the $ \nu = 0 $ state.\cite{ph2clu:14Nabi}

We can now express the many particle potential as a sum of the aforementioned pair potentials. For $ \nu_t = 0 $, this potential is
\begin{equation} \label{eq:manybody_potential_nu0}
	U^{0}(\vb{x}_1, ..., \vb{x}_N) = \sum_{i < j} V^{0}(\left| \vb{x}_i - \vb{x}_j \right|) .
\end{equation}
The expression for the excited potential surface is more complicated. Our formalism only allows the placement of the vibrational quantum at a specific site, and to account for this limitation is a non-trivial combinatorics problem. Let $ ^{N}U^{1}_{n} $ be the excited potential energy for a system of $ N $ particles, where the vibrational excitation is located on particle $ n $. For $ N = 2 $ the solution is trivial - we simply recover the excited pairwise curve 
\begin{equation} \label{eq:excited2}
^{2}U^{1}_{1}(\vb{x}_1, \vb{x}_2) = {}^{2}U^{1}_{2}(\vb{x}_1, \vb{x}_2) = V^1(R_{12}) .
\end{equation}
\noindent
For $ N = 3 $, there are three possible locations for the excitation,
\begin{align}
	{}^{3}U^{1}_{1}(\vb{x}_1, \vb{x}_2, \vb{x}_3) &= V^1(R_{12}) + V^1(R_{13}) + V^0(R_{23}) \nonumber \\
	{}^{3}U^{1}_{2}(\vb{x}_1, \vb{x}_2, \vb{x}_3) &= V^1(R_{12}) + V^0(R_{13}) + V^1(R_{23}) \nonumber \\
	{}^{3}U^{1}_{3}(\vb{x}_1, \vb{x}_2, \vb{x}_3) &= V^0(R_{12}) + V^1(R_{13}) + V^1(R_{23}) . \label{eq:excited3}
\end{align}
\noindent
Notice that the symmetric combination of the three expressions above, $ ^{3}U^{1} = \frac{1}{3} ( {}^{3}U^{1}_{1} + {}^{3}U^{1}_{2} + {}^{3}U^{1}_{3} ) $, has repeated terms that need to be dealt with. For a general $ N $, the excited pairwise potential can be expressed as
\begin{equation} \label{eq:excitedN}
	{}^{N}U^{1}(\vb{x}_1, ..., \vb{x}_N) = c^0 \sum_{i < j} V^{0}(R_{ij}) + c^1 \sum_{i < j} V^{1}(R_{ij}) ,
\end{equation}
\noindent
where $ c^0 $ and $ c^1 $ are the fraction of terms in the ground and excited state sums respectively, and $ c^0 + c^1 = 1 $. We see by examining Eq.~(\ref{eq:excited3}) that for $ N = 3 $, the symmetric combination yields doubles of each excited term. This is also true for $ N \geq 4 $, because the term $ V^{1}(R_{ij}) $ appears in both $ {}^{N}U^{1}_{i} $ and $ {}^{N}U^{1}_{j} $. Thus we conclude that $ c^1 = 2/N $, and by extension, $ c^0 = (N-2)/N $. The final expressions for the ground state and excited state Hamiltonians for a (\para)$_N$ cluster can be written, respectively, as
\begin{equation} \label{eq:HAM0}
	\hat{H}^0(\vb{x}_1, \vb{p}_1, ..., \vb{x}_N, \vb{p}_N) = \frac{1}{2 \mu} \sum_{i=1}^N \vb{p}_i^2 + \sum_{i < j} V^0(R_{ij}) ,
\end{equation}
\noindent
and
\begin{multline} \label{eq:HAM1} 
	\hat{H}^1(\vb{x}_1, \vb{p}_1, ..., \vb{x}_N, \vb{p}_N) = \frac{1}{2 \mu} \sum_{i=1}^N \vb{p}_i^2 \\ + \frac{N-2}{N} \sum_{i < j} V^0(R_{ij}) + \frac{2}{N} \sum_{i < j} V^1(R_{ij}) .
\end{multline}

\subsection{Path Integral Monte Carlo method} \label{sub:pimc}

Given a temperature $ \beta = 1/k_B T $ and the density operator $ e^{-\beta \hat{H}} $, the canonical average of a quantum mechanical operator $\hat{O}$ is expressed using the PIMC method as
\begin{align} \label{eq:DMAVG}
	\expval{\hat{O}}_{\beta}
	&= \frac{1}{Z} \mathrm{Tr} \left\{ \hat{O} e^{- \beta \hat{H}} \right\} \nonumber \\
	&= \frac{1}{Z} \int \dd \vb{q} \dd \vb{q}' \mel{\vb{q}}{\hat{O}}{\vb{q}'} \mel{\vb{q}'}{ e^{- \beta \hat{H}} }{\vb{q}} ,
\end{align}
\noindent
where $ Z = \mathrm{Tr} \{ e^{- \beta \hat{H}} \} $ is the partition function. The translational coordinates of the \para molecules are denoted collectively as $ \vb{q} = (\vb{r}_0, \{ \vb{r}_{i > 0} \} ) $. A modified version of the finite temperature MoRiBS-PIMC code developed by T. Zeng \etal\cite{pi:16tobyb}~was used to compute the density matrix and obtain the expectation value of the operator $ \hat{O} $. The integral is discretized in imaginary time,
\begin{equation} \label{eq:DMELE}
	\mel{\vb{q}'}{e^{-\beta \hat{H}}}{\vb{q}} = \int \prod^{P}_{i = 2} \dd \vb{q}_i \times \prod^{P}_{i = 1} \mel{\vb{q}_i}{e^{-\tau \hat{H}}}{\vb{q}_{i+1}} ,
\end{equation}
\noindent
where $ \tau = \beta / P $ is the imaginary time step\cite{pi:95cepe}~and the individual time steps are labeled by $ i = 1 $ to $ P $. In addition, the boundary conditions $ \vb{q}_1 = \vb{q}' $ and $ \vb{q}_{P + 1} = \vb{q} $ are applied to Eq.~(\ref{eq:DMELE}). Bose exchange is practically absent in the density ranges explored and is not included in the simulations - its inclusion barely changes the calculated quantities while increasing the amount of computation time by roughly $ 40 $ \%.

The vibrational transition frequency of the chromophore molecule changes when embedded in the \para solid. Define $ \Delta E_{\rm embed} $ and $ \Delta E_{\rm free} $ to be the transition frequencies of an embedded and free molecule, respectively. The shift of the vibrational band origin of the dopant in the \para solid relative to that of the free chromophore molecule is
\begin{equation} \label{eq:old_shift}
	\Delta \nu = \Delta E_{\rm embed} - \Delta E_{\rm free} .
\end{equation}
\noindent
We can also evaluate Eq.~(\ref{eq:old_shift}) as the difference between $ E_{\rm matrix}^{\nu_{\rm t} = 1} $, the total energy of a \para lattice in which one \para molecule is in its $ \nu = 1 $ state while all the others are in the $ \nu = 0 $ state, and $ E_{\rm matrix}^{\nu_{\rm t} = 0} $, the total energy in which all \para molecules are in the $ \nu = 0 $ state,
\begin{equation} \label{eq:old_cal_shift}
	\Delta \nu = E_{\rm matrix}^{\nu_t = 1} - E_{\rm matrix}^{\nu_t = 0} .
\end{equation}
\noindent
In other words, in the pure \para solid, the chromophore molecule is \para itself. Just as in pure \para clusters,\cite{ph2clu:14Nabi}~the ``chromophore'' \para molecule in the solid cannot be singled out from the other \para molecules in realistic simulations, and the excitation $ \nu_{\rm t} = 1 $ can diffuse by exchange.

The Hamiltonian of \spara in the excited state $ \nu_{\rm t} = 1 $ can be approximated\cite{ph2clu:14Nabi} as
\begin{equation} \label{eq:EXHA}
	\hat{H}^{1} = \hat{H}^{0} + \Delta \hat{V} ,
\end{equation}
\noindent
where $ \Delta \hat{V} = \hat{V}^{1} - \hat{V}^{0} $ is the difference between the total potential energy of \spara in the $ \nu_{\rm t} = 1 $ and $ \nu_{\rm t} = 0 $ states. Using first-order perturbation theory,\cite{dhe:09hlia, dhe:12wang, ph2clu:14Nabi}~we can write the vibrational frequency shift as
\begin{equation}
	\Delta \nu = \expval{\Delta \hat{V}} .
\end{equation}
\noindent
This quantity is determined by the radial distribution function $ g(r) $ of the \para molecules in the solid. However, it is easier to directly retrieve from the PIMC simulation the weight distribution function $ h(r) $, where $ h(r) \, \dd r = 4 \pi r^2 \rho \, g(r) \, \dd r $, which is then normalized such that $ \int \dd r \, h(r) = 1 $. We can evaluate the vibrational frequency shift in a simulation of $ N $ \para molecules as
\begin{align} \label{eq:new_shift}
	\Delta \nu
	&= 4 \pi \rho (N - 1) \int \dd r \ r^2 g(r) \Delta V (r) , \nonumber \\
	&= (N - 1) \int \dd r \ h(r) \Delta V (r) ,
\end{align}
\noindent
where $ \Delta V(r) = V^1(r) - V^0(r) $.

The use of periodic boundary conditions allows us to simulate \para crystals of various densities for both the \fcc and \hcp structures. Each simulation is performed by constructing a system of $ \mathrm{D}_1 \times \mathrm{D}_2 \times \mathrm{D}_3 $ elementary cells for a given lattice type. The integers $ \mathrm{D}_1 $, $ \mathrm{D}_2 $, and $ \mathrm{D}_3 $ are chosen such that the ratio of the three sides of the system to each other is as close as possible to unity for cells of manageable sizes. Cells of size $ 5 \times 3 \times 3 $, $ 7 \times 4 \times 4 $, $ 8 \times 5 \times 5 $, and $ 10 \times 6 \times 6 $ were constructred for the \hcp structre, and cells of size $ 4 \times 4 \times 4 $, $ 5 \times 5 \times 5 $, $ 6 \times 6 \times 6 $, and $ 7 \times 7 \times 7 $ were used for the \fcc structure. Each elementary cell contains $ 4 $ \para molecules, for both the \hcp and \fcc structures. The energy and vibrational band shift measurements from these systems of different finite sizes are then extrapolated to obtain their counterparts for a system of infinite size. This extrapolation is discussed in detail in Secs.~\ref{sub:tauextrap} and \ref{sub:nextrap}.

\section{Numerics\label{sec:numerics}}
\subsection{Extrapolating to an infinite number of time steps} \label{sub:tauextrap}

The use of finitely many time slices in calculations is the primary source of systematic errors in any PIMC simulation. To compensate for this error, the calculation can be repeated several times for different numbers of time slices. We can then use the expression
\begin{equation} \label{eq:tauextrap}
	\expval{\hat{O}}_{N, \tau} = \expval{\hat{O}}_{N, 0} + B_2 \, \tau^2 + B_4 \, \tau^4
\end{equation}
\noindent
to extrapolate the PIMC data to the limit $ \tau \rightarrow 0 $.\cite{pimc:17yan}~Here, $ \langle \hat{O} \rangle _{N, \tau} $ can be either the energy $ E $ or the band origin shift $ \Delta \nu $ for a system of $ N $ \para molecules simulated with the time step size $ \tau $. The parameters $ \langle \hat{O} \rangle _{N,0} $, $ B_2 $ and $ B_4 $ are determined by applying Eq.~(\ref{eq:tauextrap}) to several $ ( \langle \hat{O} \rangle _{N, \tau}, \tau ) $ pairs, and $ \langle \hat{O} \rangle _{N, 0} $ is the extrapolated value.
\begin{figure} [H]
	\centering
	\includegraphics[width=\linewidth]{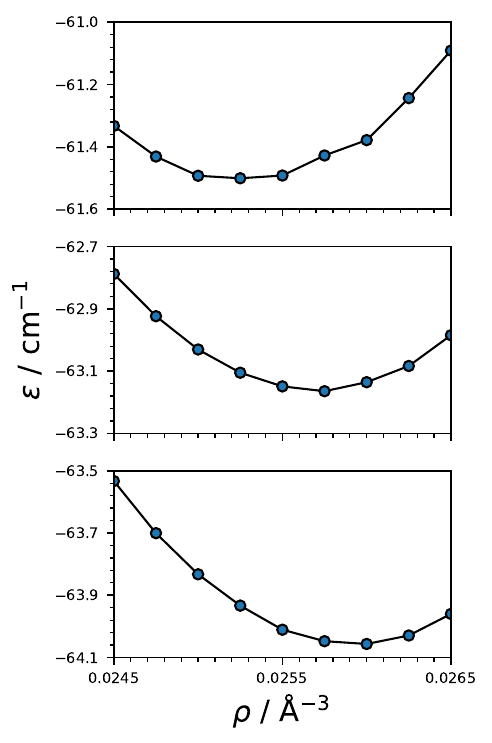}
	\caption{Energy per particle $ \epsilon $ as a function of density $ \rho $ obtained for numbers of time steps $ P = 192 $ (top), $ P = 96 $ (centre), and $ P = 80 $ (bottom) in the PIMC simulation, where $\tau=\beta/P$. Each simulation was performed using $ N = 180 $ \para molecules. The density for which the EOS is at its minimum changes with $ P $.}
	\label{fig:paper0_fig1}
\end{figure}

\begin{figure} [H]
	\centering
	\includegraphics[width=\linewidth]{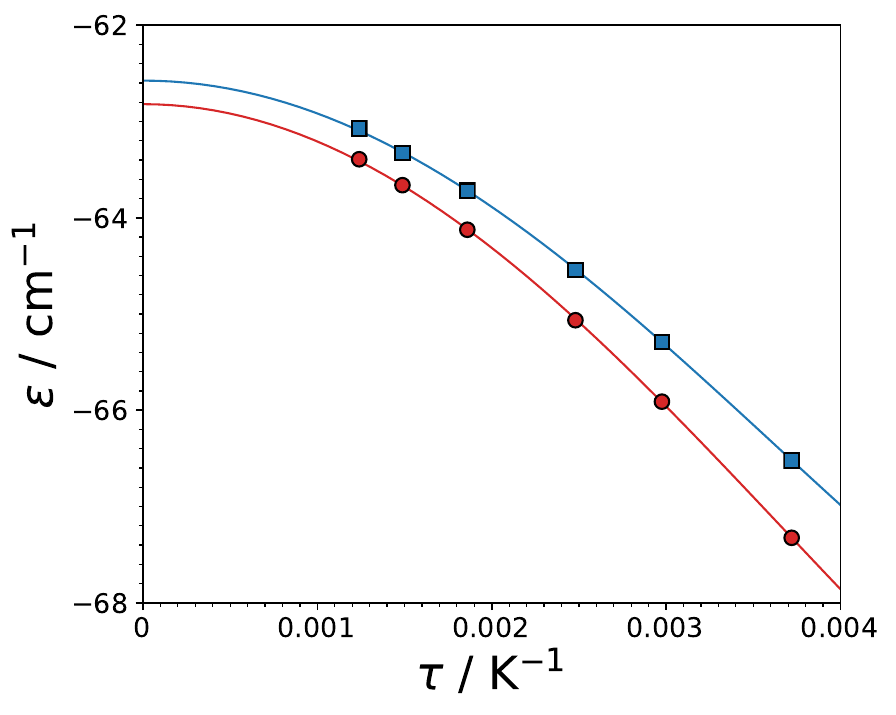}
	\caption{Energy per particle $ \epsilon $ as a function of imaginary time step size $ \tau $ of an \hcp crystal with $ N = 1440 $ \para molecules and a density of $ \rho = 0.0255 \, \ANG^{-3} $ (blue squares) and $ \rho = 0.0245 \, \ANG^{-3} $ (red circles). The value of $ \epsilon $ at $ \tau = 0 $ is found using Eq.~(\ref{eq:tauextrap}).}
	\label{fig:paper0_fig2}
\end{figure}

Most previously reported simulations of doped rare gas clusters\cite{dhe:11wang, dhe:12wang}~started with extensive coverage tests to choose a value of $ \tau $ that is ``small enough'' such that $ \langle \hat{O} \rangle_{N, \tau} $ no longer changes appreciably by decreasing $ \tau $. The working calculations were then performed using only the chosen $ \tau $. To see why the extrapolation $ \tau \rightarrow 0 $ is important, we show the energy per particle $ \epsilon $ for an \hcp crystal of $ N = 180 $ \para molecules for different numbers of time slices in Fig.~(\ref{fig:paper0_fig1}). One calculation of interest is to determine the equilibrium density of \spara, and we see in Fig.~(\ref{fig:paper0_fig1}) that the equilibrium density of the EOS curve changes with $ \tau $. In Fig.~(\ref{fig:paper0_fig2}) the energy per particle is plotted as a function of the imaginary time step for two densities, and fit using Eq.~(\ref{eq:tauextrap}). For both examples, the extrapolated value is roughly $ 0.5 $ cm$^{-1}$ greater than the value found using $ P = 192 $ time slices. We will see in Sec.~\ref{sub:equilibriumEOS} that this correction is an order of magnitude greater than the difference between the ground state energy per particle of the \hcp and \fcc crystals. It is therefore necessary to extrapolate $ \tau $ to the limit $ \tau \rightarrow 0 $ rather than to rely on a ``small enough $ \tau $'' approximation.

Because most observed and predicted equations of state were obtained at $ T = 4.2 $~K,\cite{ph2solex:63good, ph2solex:75dura, ph2solth:78silv, ph2solex:79drie, ph2solex:80silv} this is the temperature used for most of our simulations. We perform simulations with the numbers of time slices $ P = \{ 64, 80, 96, 128, 160, 192 \} $, and then use Eq.~(\ref{eq:tauextrap}) to extrapolate the energy and band origin shift measurements to $ \tau \rightarrow 0 $. To study the temperature dependence of the EOS and vibrational matrix shift, an \hcp periodic cell of $ N = 448 $ \para molecules is simulated at temperatures $ T = \{ 2.10 \, \mathrm{K}, 4.20 \, \mathrm{K}, 5.04 \, \mathrm{K}, 6.30 \, \mathrm{K} \} $, with the number of time slices adjusted such that $ \tau $ is the same for each simulation.

\subsection{Extrapolating to an infinite number of particles} \label{sub:nextrap}

The use of a finite number of \para molecules introduces an additional source of systematic error to the periodic boundary condition methodology, called the finite size effect.\cite{pbc:04yeh, pbc:09naka}~To account for this error we calculate physical observables for systems of different sizes and extrapolate these measurements to $ N \rightarrow \infty $.

\begin{figure} [t]
	\centering
	\includegraphics[width=\linewidth]{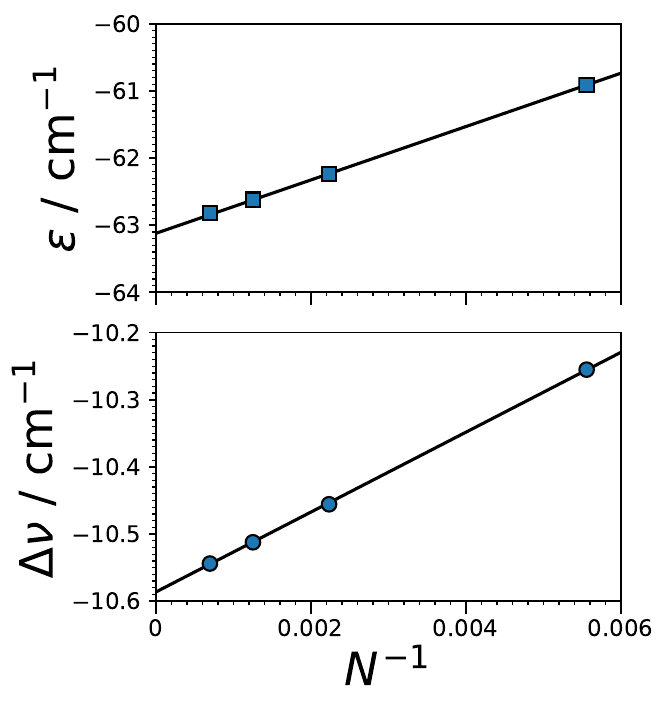}
	\caption{The energy per particle ($ \epsilon $, top) and vibrational frequency shift ($ \Delta \nu $, bottom) at $ \tau = 0 $ as a function of the number of \para molecules $ N $, for an \hcp crystal at density of $ \rho = 0.0255 \, \ANG^{-3} $. Using Eq.~(\ref{eq:epsilon_contribution}), we find both these quantities for an infinite lattice of \para molecules.}
	\label{fig:paper0_fig3}
\end{figure}

In Fig.~(\ref{fig:paper0_fig3}) we see that, after having performed the imaginary time extrapolation $ \tau \rightarrow \infty $, both the the energy per particle $ \epsilon = E / N $ and the band origin shift $ \Delta \nu $ are linear functions of $ 1/N $. To explain why, consider a periodic cell with an edge length of $ 2 L $. Due it its finite size, there are additional \para molecules located beyond a shell of radius $ r > L $, where $ r $ is the distance between the centre of the periodic cell and the surface of the shell. At long ranges, the function $ V^0(r) $ is dominated by the leading term of the dispersion interaction, which is proportional to $ r^{-6} $.\cite{h2pes:10hli}~Also, the weight distribution function $ h(r) $ is determined by the density of \para on the surface area of the shell, so we can write $ h(r) \sim r^2 $. As given by the tail correction,\cite{tail:87allen} the contribution to $ \epsilon $ from the additional \para molecules beyond the shell is
\begin{equation} \label{eq:epsilon_contribution}
	\Delta \epsilon = 2 \pi \rho \int \dd r \, h(r) V^0(r) \sim \int_{L}^{\infty} \dd r \, r^2 \frac{1}{r^6} \sim \frac{1}{L^3} .
\end{equation}
\noindent
Because the density is uniform, the volume of the periodic cell is proportional to the number of \para molecules within, and we can write $ \Delta \epsilon \sim L^{-3} \sim N^{-1} $. A similar calculation shows that the contribution from the additional \para molecules to the vibrational matrix shift is also inversely proportional to $ N $. Thus to account for the finite size effect, simulations are performed for different numbers of particles and the PIMC data is extrapolated to the limit $ N \rightarrow \infty $ using
\begin{equation} \label{eq:nextrap}
	\expval{\hat{O}}_{N, 0} = \expval{\hat{O}}_{\infty, 0} + \frac{C}{N} .
\end{equation}
\noindent
In Eq.~(\ref{eq:nextrap}), $ \langle \hat{O} \rangle _{N, 0} $ is the observable calculated using a system of $ N $ \para molecules with the extrapolation in Eq.~(\ref{eq:tauextrap}) already applied, $ \langle \hat{O} \rangle_{\infty, 0} $ is its counterpart with the extrapolation $ N \rightarrow \infty $ applied, and $ C $ is a constant. The values of $ \epsilon $ and $ \Delta \nu $ obtained with Eq.~(\ref{eq:nextrap}) are lower because more \para--\para interaction pairs are included in a region where the potential is attractive.

We see from Eq.~(\ref{eq:epsilon_contribution}) that the scaling of $ \epsilon $ and $ \Delta \nu $ with $ N $ is determined by the leading term of the long-range interaction - if the leading term is proportional to $ r^{-m} $, then $ \epsilon $ and $ \Delta \nu $ change linearly with $ N^{1 - m/3} $.

\section{Results and Discussion\label{sec:results}}
\subsection{The global EOS} \label{sub:globalEOS}

\begin{figure} [H]
	\centering
	\includegraphics[width=\linewidth]{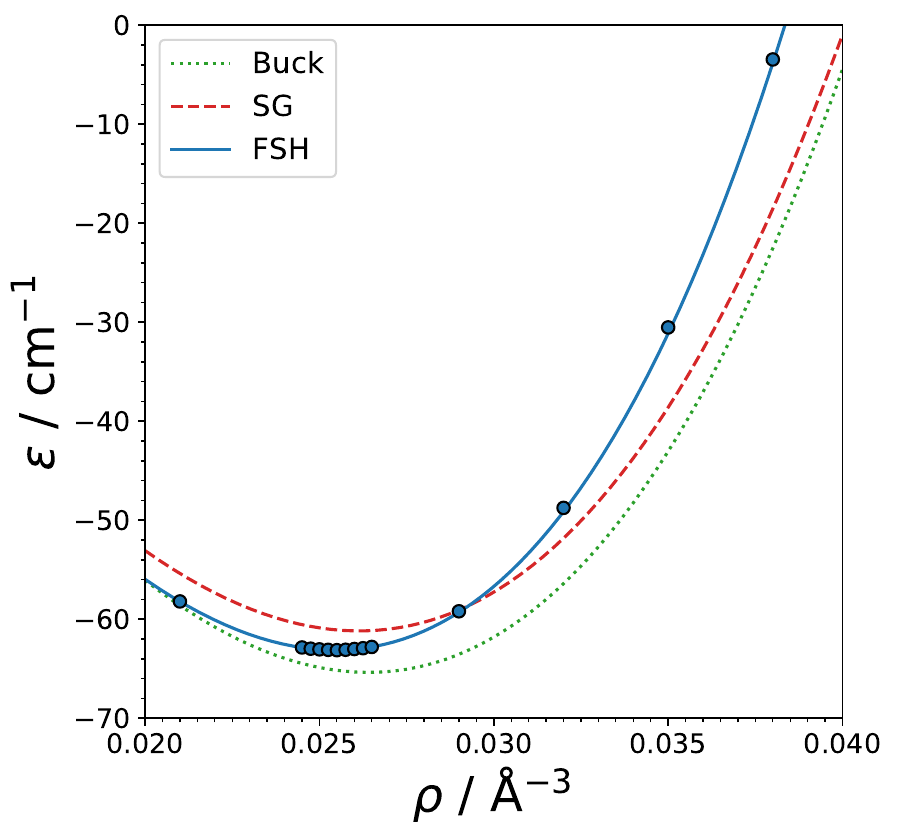}
	\caption{The EOS curves for an \hcp lattice of \para molecules using the FSH potential at $ T = 4.2 $~K (solid blue, circles), the Buck potential at $ T = 0 $~K (green, dotted), and the SG potential at $ T = 0 $~K (red, dashed). The FSH potential provides the lowest equilibrium density of the three.}
	\label{fig:paper0_fig4}
\end{figure}

To fit the calculated values of the energy per particle we use a Murnaghan-type curve\cite{fiteq:44murn}
\begin{equation} \label{eq:murnaghan}
	\epsilon(\rho) = a + b \rho + c \rho^{\gamma} ,
\end{equation}
\noindent
where $ \epsilon $ is the energy per particle, $ \rho $ is the density, and $ (a, b, c, \gamma) $ are the fit parameters. The EOS for an \hcp crystal, after having performed the $ \tau \rightarrow 0 $ and $ N \rightarrow \infty $ extrapolations, is presented in Fig.~(\ref{fig:paper0_fig4}) on the density interval $ 0.02 \, \ANG^{-3} < \rho < 0.04 \, \ANG^{-3} $. Alongside it we also include several previously reported EOS curves.\cite{ph2solth:06Oper} We can see from Fig.~(\ref{fig:paper0_fig4}) that the resulting curve fits the measured data satisfyingly. 

\begin{table*}[ht]
	\centering
	\caption{Coefficients $a$, $b$, $c$, and $\gamma$ obtained by fitting Eq.~(\ref{eq:murnaghan}) to the ($ \rho $, $ \epsilon $) pairs obtained from our PIMC simulation and from Ref.~\citenum{ph2solth:06Oper}. The density $ \rho $ is given in \AA$^{-3}$, and the energy per particle $\epsilon$ is given in cm$^{-1}$. Note that our results corrected for the finite size effect using Eq.~(\ref{eq:nextrap}) while in Ref.~\citenum{ph2solth:06Oper} the tail correction was performed.}
	\setlength{\tabcolsep}{.5em}

    \begin{tabular}{l@{\hskip 0.15in}l@{\hskip 0.15in}l@{\hskip 0.15in}l@{\hskip 0.15in}l}
        \hline\hline
        ~                                                      & \thead{$a$} & \thead{$b$} & \thead{$c$} & \thead{$\gamma$}  \\ \hline
        FSH(\fcc, $T = 4.2$~K) & $28.1712$   & $-4791.96$  & $6.32311 \times 10^7$ & $3.96$    \\ 
        FSH(\hcp, $T = 4.2$~K) & $27.5055$   & $-4762.14$  & $6.29562 \times 10^7$ & $3.96$    \\ 
        Buck(\hcp, $T = 0$~K)                       & $36.4411$   & $-5380.89$  & $1.49110 \times 10^7$ & $3.52809$ \\ 
        SG(\hcp, $T = 0$~K)            & $33.4464$   & $-5058.51$  & $1.38902 \times 10^7$ & $3.51769$ \\
        \hline\hline
    \end{tabular}
	
	\label{tab:EOS_params}
\end{table*}

\begin{figure} [H]
	\centering
	\includegraphics[width=\linewidth]{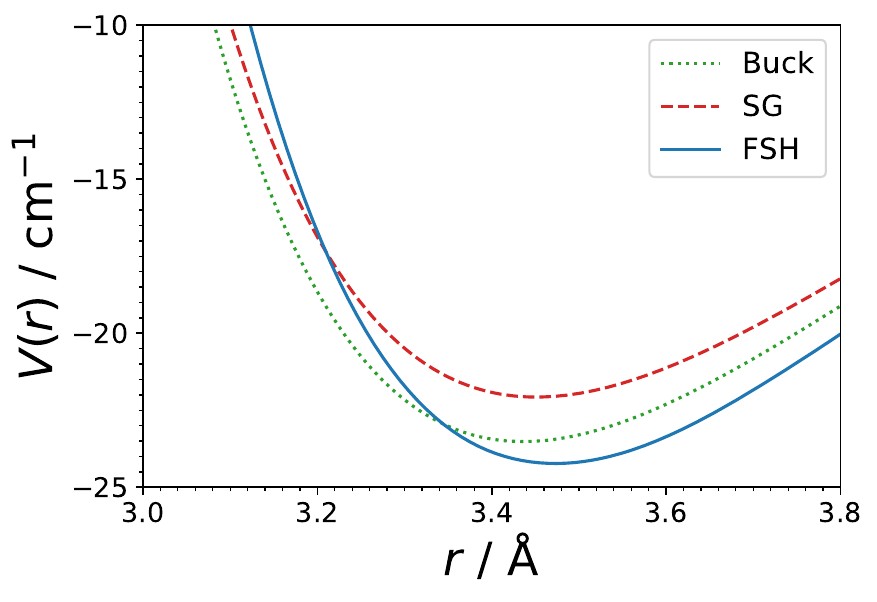}
	\caption{The \para--\para potential energy curves for the FSH potential (blue, solid), the Buck potential (green, dotted) and the SG potential (red, dashed). The FSH potential has the greatest repulsive wall when $ R < 3.2 \ \ANG $, and so it predicts a lower equilibrium density than the other two potentials.}
	\label{fig:paper0_fig5}
\end{figure}

On this density interval, the EOS's for \fcc and \hcp crystals are nearly indistinguishable, and so the former is omitted.\cite{ph2solth:06Oper} The fit parameters $ a $, $ b $, $ c $, and $ \gamma $ are provided for the curves in Fig.~(\ref{fig:paper0_fig4}) in Tab.~\ref{tab:EOS_params}. From this fit we can determine the equilibrium density $ \rho_0 $, the distance between two nearest neighbours $ R_0 $, and the equilibrium energy per particle $ \epsilon_0 $, which we summarize in Tab.~\ref{tab:EOS_properties}. The nature of these EOS curves is closely related to the position and depth of their corresponding potential energy curves, shown in Fig.~(\ref{fig:paper0_fig5}). For example, we can see in Fig.~(\ref{fig:paper0_fig4}) that the equilibrium density of the EOS curve of the FSH potential is lower than those of the SG and Buck potentials. This is because its minimum is deeper and lies at a greater distance that those of the other potentials. The FSH potential also has the greatest repulsive wall of the three, which is why, at high densities, its EOS curve is greater than those of the SG and Buck potentials. In the same fashion, the Buck potential has the lowest repulsive wall, which is consistent with the observation that its EOS lies below the others at high densities.

\begin{table*} [ht]
	\centering
	\caption{The equilibrium density $ \rho_0 $ (in $\ANG^{-3}$), the distance between nearest neighbours $ R_0 $ (in \ANG), and the equilibrium energy per particle $ \epsilon_0 $ (in cm$^{-1}$), of \fcc and \hcp \para crystals, obtained in our PIMC simulations and from Ref.~\citenum{ph2solth:06Oper}. Also included are experimental results from Refs.~\citenum{ph2solex:80silv} and \citenum{ph2solth:06Oper}. Note that our results corrected for the finite size effect using Eq.~(\ref{eq:nextrap}) while in Ref.~\citenum{ph2solth:06Oper} the tail correction was performed.}

    \begin{tabular}{@{\extracolsep{6pt}}l@{\hskip 0.08in}l@{\hskip 0.08in}l@{\hskip 0.08in}l@{\hskip 0.08in}l@{\hskip 0.08in}l@{\hskip 0.08in}l}
        \hline\hline
		~                                                & \multicolumn{3}{c}{\fcc} & \multicolumn{3}{c}{\hcp} \\
		\cline{2-4}\cline{5-7}
        ~                                                & \thead{$\rho_0$} & \thead{$R_0$} & \thead{$\epsilon_0$} & \thead{$\rho_0$} & \thead{$R_0$} & \thead{$\epsilon_0$} \\ \hline
        FSH($T = 4.2$~K) & $0.02547$ & $3.815$ & $-63.062$ & $0.02546$ & $3.816$ & $-63.102$ \\ 
        Buck($T = 0$~K)                       & $0.02647$ & $3.766$ & $-64.99$ & $0.02640$ & $3.770$ & $-65.36$ \\ 
        SG($T = 0$~K)            & $0.02606$ & $3.786$ & $-60.82$ & $0.02613$ & $3.783$ & $-61.18$ \\
		exp($ T = 0 $~K)                                 & \multicolumn{1}{c}{-} & \multicolumn{1}{c}{-} & \multicolumn{1}{c}{-} & $0.02600$\cite{ph2solex:80silv} & $3.789$ & $-62.5$\cite{ph2solth:06Oper} \\
        \hline\hline
    \end{tabular}
	
	\label{tab:EOS_properties}
\end{table*}

We should also note that our simulations were performed at $ T = 4.2 $~K with the results extrapolated to $ N \rightarrow \infty $. The DMC simulations for the Buck and SG potentials were performed at absolute zero and for a system of $ N = 108 $ \para molecules with a tail correction applied.\cite{ph2solth:06Oper}

\subsection{EOS near equilibrium density} \label{sub:equilibriumEOS}

\begin{figure} [H]
	\centering
	\includegraphics[width=\linewidth]{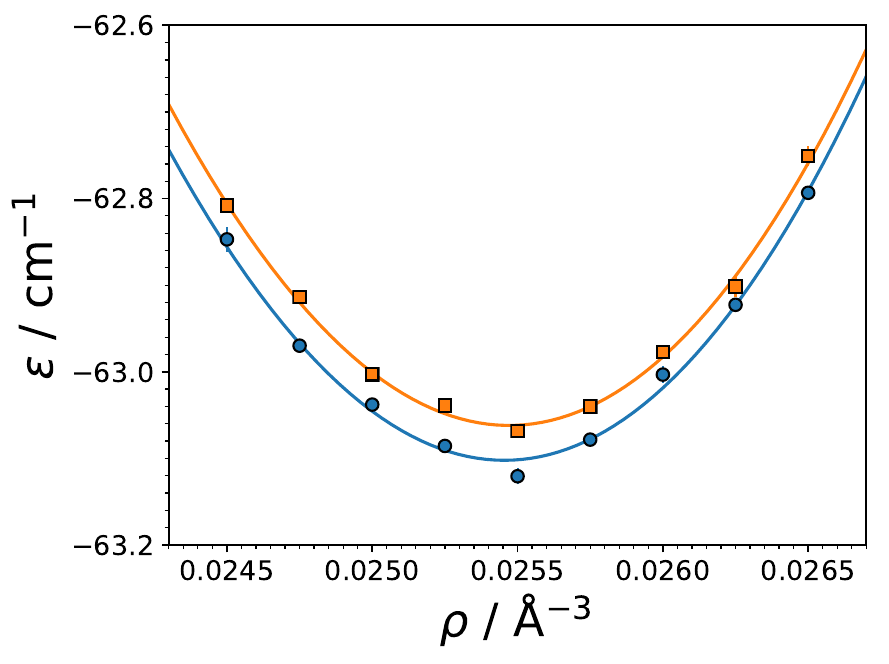}
	\caption{The EOS of \spara near the equilibrium density for the \hcp crystal (blue circles) and the \fcc crystal (orange squares) after performing the imaginary time step extrapolation $ \tau \rightarrow 0 $ and the system size extrapolation $ N \rightarrow \infty $. The \hcp lattice structure is the more stable of the two.}
	\label{fig:paper0_fig6}
\end{figure}

The EOS curves of \spara in both its \fcc and \hcp forms near the equilibrium density are shown in Fig.~(\ref{fig:paper0_fig6}). All the $ ( \rho, \epsilon ) $ points here were obtained by extrapolating both $ \tau $ and $ N $. We can see that the values of $ \epsilon $ for the \hcp crystal are always lower than those of the \fcc crystal by about $ 0.04 $ cm$^{-1}$, making the \hcp crystal the more stable of the two. This result agrees with experiments indicating that the \hcp lattice is the preferred structure of a \para crystal around $ T = 4.2 $~K.\cite{ph2solex:68schu,ph2solex:80silv}

\begin{figure} [t]
	\centering
	\includegraphics[width=\linewidth]{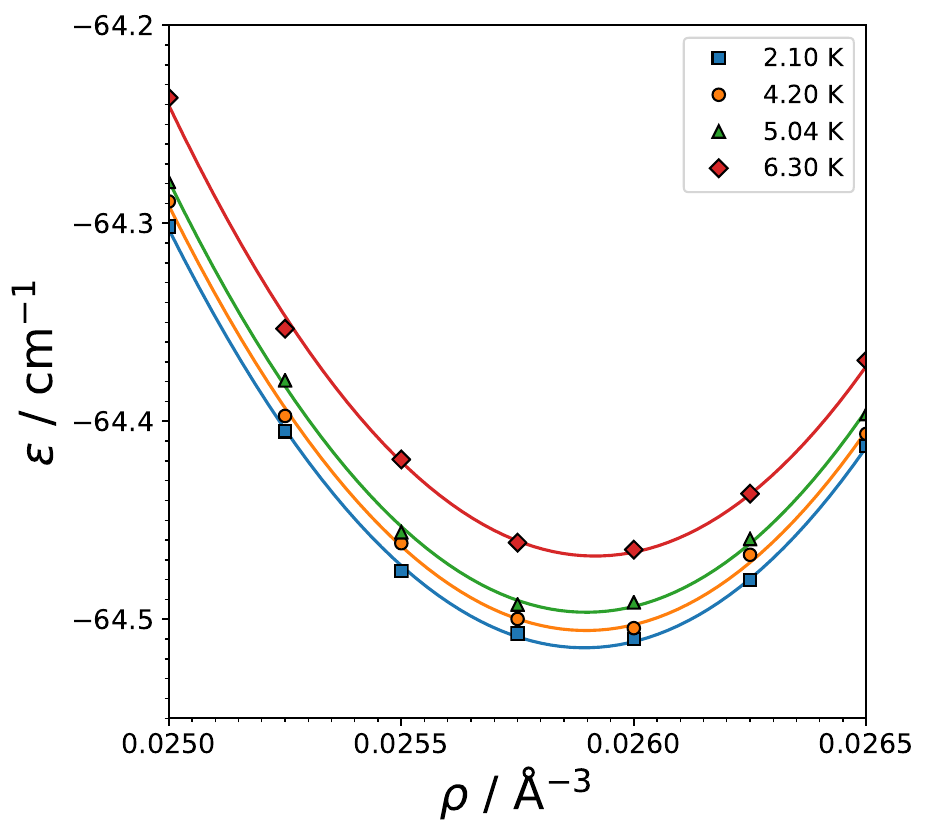}
	\caption{The EOS for a \para \hcp crystal at temperatures of $ T = 2.10 $~K (blue squares), $ T = 4.20 $~K (orange circles), $ T = 5.04 $~K (green triangles), and $ T = 6.30 $~K (red diamonds), with $ N = 448 $ and $ P $ adjusted such that in each simulation, $ \tau \approx 0.00248 $~K$^{-1}$. The energy per particle at equilibrium decreases with temperature, but the equilibrium density itself remains unchanged.}
	\label{fig:paper0_fig7}
\end{figure}

To study the temperature dependence of the EOS, we simulate the $ 7 \times 4 \times 4 $ \hcp lattice at temperatures $ T = \{ 2.10 \, \mathrm{K}, 4.20 \, \mathrm{K}, 5.04 \, \mathrm{K}, 6.30 \, \mathrm{K} \} $, with the number of time slices adjusted such that $ \tau \approx 0.00248 $~K$^{-1}$ for each simulation. In other words, no $ N $ or $ \tau $ extrapolation was performed for these simulations. The results in Tab.~\ref{tab:EOS_temperature_properties} and Fig.~(\ref{fig:paper0_fig7}) indicate that as the temperature decreases, the EOS isotherm becomes deeper and the value of $ \epsilon_0 $ decreases. However, the changes in the EOS isothem become less substational at lower temperatures. For example, the change in the EOS curve from $ T = 4.20 $~K to $ T = 2.10 $~K is much smaller than the change from $ T = 6.30 $~K to $ T = 5.04 $~K, despite being a larger decrease in temperature. These observations agree with the general temperature invariance of the EOS curve of \spara below $ T = 4.2 $~K, mentioned in Ref.~\citenum{ph2solex:80silv}.

\begin{table} [ht]
	\centering
	\caption{The equilibrium density $ \rho_0 $ (in $\ANG^{-3}$), the distance between nearest neighbors $ R_0 $ (in \ANG), and the equilibrium energy per particle $ \epsilon_0 $ (in cm$^{-1}$), of \hcp \para crystals at different temperatures. All simulations here are performed using the FSH potential, with $ \tau \approx 0.00254 $~K$^{-1}$ and $ N = 448 $ \para molecules. The values are obtained using the fit Eq.~(\ref{eq:murnaghan}), while fixing $ \gamma = 3.96 $.}

    \begin{tabular}{l@{\hskip 0.15in}l@{\hskip 0.15in}l@{\hskip 0.15in}l}
        \hline\hline
        ~              & \thead{$\rho_0$} & \thead{$R_0$} & \thead{$\epsilon_0$} \\ \hline
        $ T = 2.10 $~K & $0.02589$        & $3.794$       & $-64.513$             \\ 
        $ T = 4.20 $~K & $0.02590$        & $3.794$       & $-64.506$             \\ 
        $ T = 5.04 $~K & $0.02590$        & $3.794$       & $-64.496$             \\ 
        $ T = 6.30 $~K & $0.02592$        & $3.792$       & $-64.468$             \\ 
        \hline\hline
    \end{tabular}
	
	\label{tab:EOS_temperature_properties}
\end{table}

\subsection{Pressure as a function of density} \label{sub:pressure}

\begin{figure} [t]
	\centering
	\includegraphics[width=\linewidth]{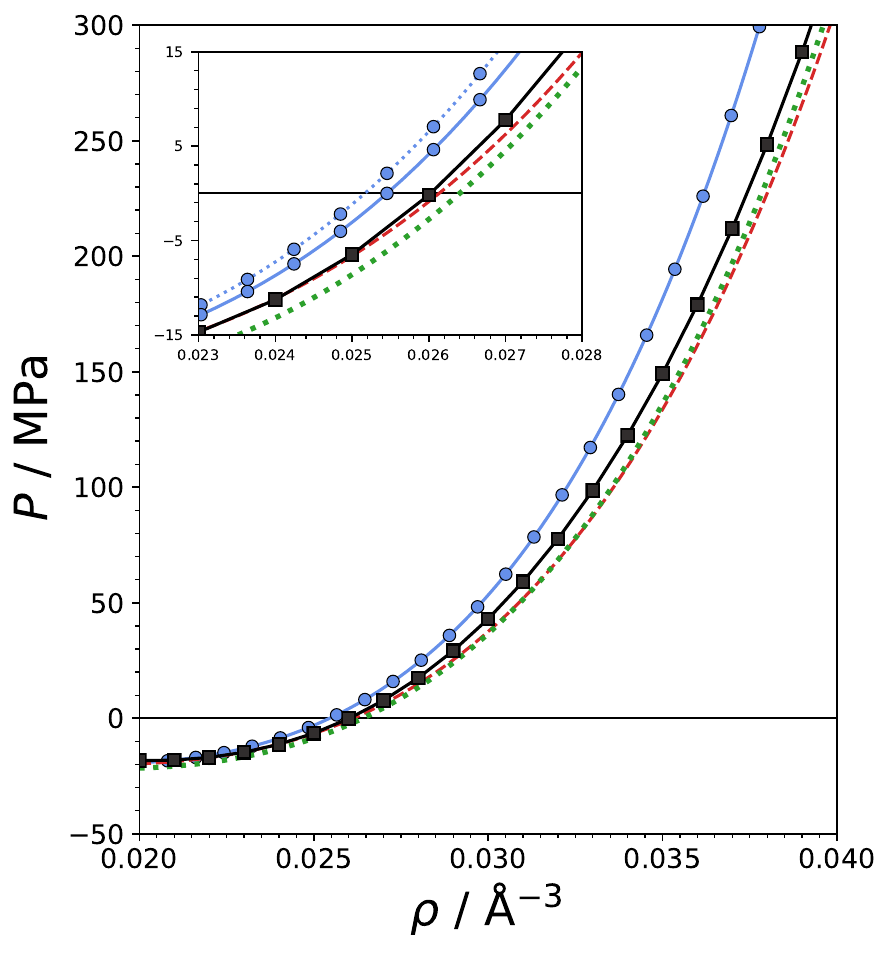}
	\caption{The pressure for an \hcp crystal of \para molecules as a function of density. Plotted are the FSH potential at $ T = 4.2 $~K extrapolated to $ N \rightarrow \infty $, (solid blue line with circles), the Buck potential at $ T = 0 $~K (dotted green line), the SG potential at $ T = 0 $~K (dashed red line) and the experimental results at $ T = 4.2 $~K (solid black line with squares). Also included within the inset is the FSH potential at $ T = 4.2 $~K and $ N = 180 $ (dotted blue line with circles).}
	\label{fig:paper0_fig8}
\end{figure}

The pressure as a function of density can be calculated from the EOS using\cite{ph2solth:06Oper}
\begin{equation} \label{eq:pressure_density}
	P = \rho^2 \pdv{\epsilon}{\rho} \eval_{T}.
\end{equation}
\noindent
In Fig.~(\ref{fig:paper0_fig8}) we see the computed pressure curves on the density interval $ 0.02 \ \ANG^{-3} < \rho < 0.04 \ \ANG^{-3} $ for the FSH, Buck, and SG potentials\cite{ph2solth:06Oper}~and experimental data.\cite{ph2solex:80silv, ph2solex:79drie}~The inset in Fig.~(\ref{fig:paper0_fig8}) presents this data near the equilibrium density. The difference between the pressure curves for the \hcp and \fcc crystals is negligible on both density ranges, so only the results for the former are plotted. This is expected from Tab.~\ref{tab:EOS_params}, which shows that the values of $ b $, $ c $ and $ \gamma $ for the \hcp and \fcc fits using Eq.~(\ref{eq:murnaghan}) are nearly the same.

We see from the tail correction Eq.~(\ref{eq:epsilon_contribution}) that the decrease in $ \epsilon $ from the extrapolation of $ N \rightarrow \infty $ becomes greater in magnitude at higher densities. This correction lowers the first derivative of the energy curve in Fig.~(\ref{fig:paper0_fig4}) and decreases the pressure at higher densities. The inset of Fig.~(\ref{fig:paper0_fig8}) shows that for the FSH potential, extrapolating the system size from $ N = 180 $ to $ N \rightarrow \infty $ results in better agreement with experimental observations.

The $ P $ \vs $ \rho $ curves are temperature-dependent.\cite{ph2solex:79drie} This may explain why our curve in the inset of Fig.~(\ref{fig:paper0_fig8}) lies above those obtained by Operetto \etal.\cite{ph2solth:06Oper} In addition, our results show greater discrepancies compared to experimental observations than do the curves obtained from Buck\cite{h2pes:83buck} and SG.\cite{h2pes:78sg} This is because our simulation was performed using a first-principles pair potential based on the interaction in a \para dimer, with neither the empirical correction in the repulsive region nor the inclusion of three-body potential terms. At short ranges, the three-body \para potential is attractive.\cite{h2h2h2pes:08hinde} Its inclusion to the FSH potential will lower the potential energy and decrease the pressure of the solid at higher densities.

\subsection{Compressibility of solid parahydrogen} \label{sub:compress}

\begin{figure} [t]
	\centering
	\includegraphics[width=\linewidth]{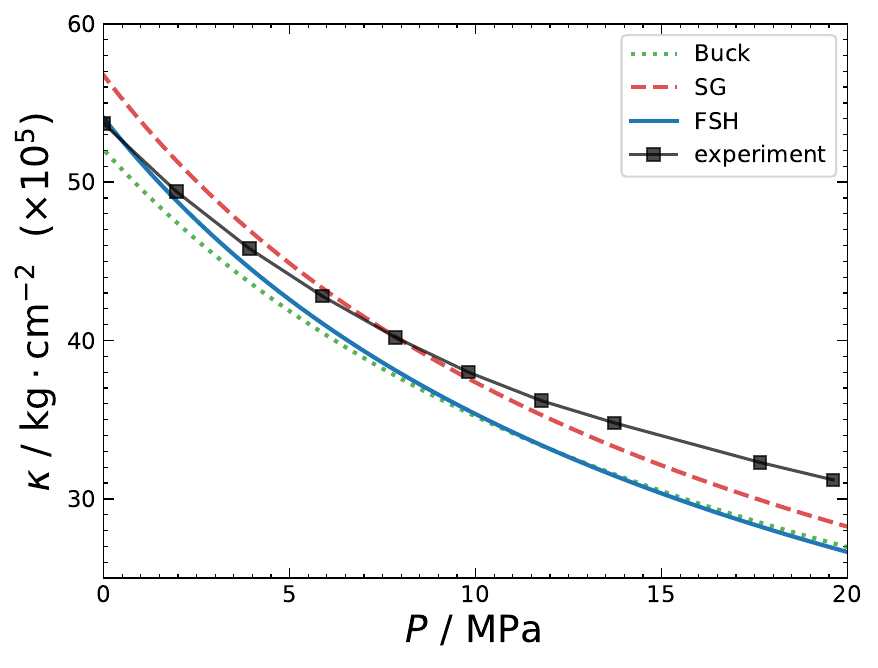}
	\caption{The compressibility $ \kappa $ for an \hcp crystal of \para molecules as a function of pressure $ P $, calculated using Eq.~(\ref{eq:compress}). Plotted are the FSH potential at $ T = 4.2 $~K (solid blue line), the Buck potential at $ T = 0 $~K (dotted green line), the SG potential at $ T = 0 $~K (dashed red line) and the experimental results for $ T < 10 $~K (solid black line with squares).}
	\label{fig:paper0_fig9}
\end{figure}

From the EOS of solid parahydrogen we can also recover its compressibility\cite{ph2solth:06Oper}, given by
\begin{equation} \label{eq:compress}
	\kappa = \frac{1}{\rho} \pdv{\rho}{P} \eval_{T} \ .
\end{equation}
\noindent
In Fig.~(\ref{fig:paper0_fig9}) we show the the compressibility as a function of pressure recovered from simulations using the FSH, Buck, and SG potentials, alongside experimental results provided in Ref.~\citenum{compr:70udov}. We see that the FSH potential is the most successful of the three at predicting the compressiblity at zero pressure, and that they each provide compressibilities that are too low as the pressure increases. In the case of the FHS potential, this deviation occurs because it is a first-principles pair interaction, and does not incorporate many-body effects. The three-body \para potential is attractive at short ranges,\cite{h2h2h2pes:08hinde} and its inclusion to a simulation alongside the FSH potential will make \spara more stable at higher densities and increase its compressibility.

\subsection{Vibrational band origin shifts} \label{sub:vibration}

\begin{figure} [h]
	\centering
	\includegraphics[width=\linewidth]{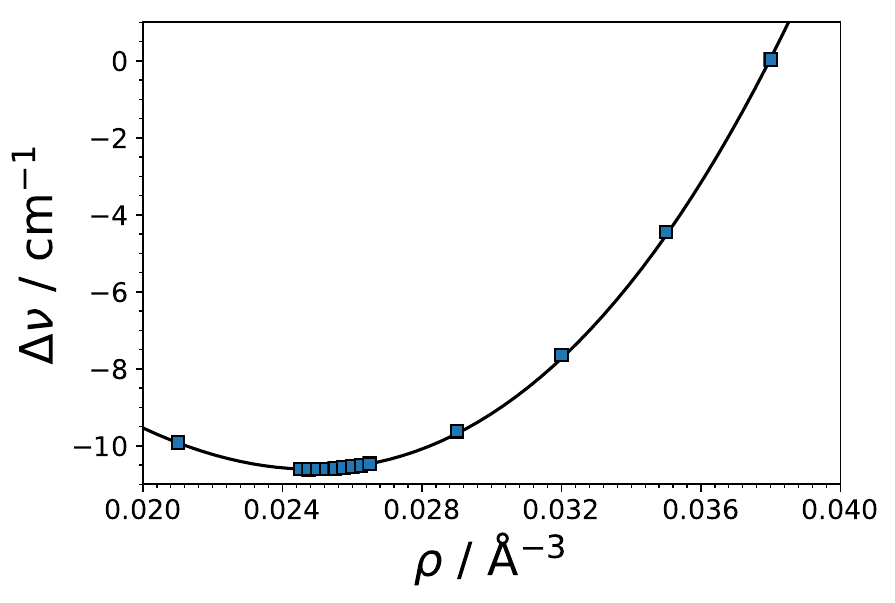}
	\caption{The vibrational matrix shift $ \Delta \nu $ for an \hcp crystal as a function of the density $ \rho $, calculated using Eq.~(\ref{eq:new_shift}) after performing the extrapolations $ \tau \rightarrow 0 $ and $ N \rightarrow \infty $. The corresponding curve for the \fcc crystal is nearly indistinguishable from that of the \hcp crystal on this density range.}
	\label{fig:paper0_fig10}
\end{figure}

The calculated vibrational matrix shift $ \Delta \nu $ as a function of density $ \rho $ for the \hcp crystal in the density interval $ 0.02 \, \ANG^{-3} < \rho < 0.04 \, \ANG^{-3} $ is shown in Fig.~(\ref{fig:paper0_fig10}). These $ \Delta \nu $ values are calculated after performing both the $ N $ and $ \tau $ extrapolation, and the resulting $ ( \rho, \Delta \nu ) $ pairs are then fit using Eq.~(\ref{eq:murnaghan}), the same equation used to fit the energy per particle. On the density range of Fig.~(\ref{fig:paper0_fig10}), the results for the \hcp and \fcc crystals are nearly indistinguishable. The minimum of the vibrational frequency shift is located near $ \rho = 0.0248 \, \ANG^{-3} $, which is noticeably lower than the equilibrium density of $ \rho_0 \approx 0.0255 \, \ANG^{-3} $ for the EOS curves.

\begin{table*} [t]
	\centering
 	\caption{Coefficients $a$, $b$, $c$, and $\gamma$ obtained by fitting Eq.~(\ref{eq:murnaghan}) to the ($ \rho $, $ \Delta \nu $) pairs obtained from our PIMC simulations. Also included is $ \Delta \nu_0 $ (in cm$^{-1}$), the vibrational matrix shift at equilibrium density $ \rho_0 $. Both the simulation and experiment results are recorded for $ T = 4.2 $~K.}

    \begin{tabular}{l@{\hskip 0.15in}l@{\hskip 0.15in}l@{\hskip 0.15in}l@{\hskip 0.15in}l@{\hskip 0.15in}l}
        \hline\hline
        ~                           & \thead{$a$} & \thead{$b$} & \thead{$c$} & \thead{$\gamma$} & \thead{$\Delta \nu_0$} \\ \hline
        \fcc, $N \rightarrow \infty$ & $8.0384$   & $-1077.01$     & $1.60252 \times 10^6$    & $3.30$         & $-10.59$               \\ 
        \hcp, $N \rightarrow \infty$ & $8.0160$   & $-1075.66$     & $1.60036 \times 10^6$    & $3.30$         & $-10.59$               \\ 
		\hcp, $N = 180$              & $7.7383$   & $-1053.37$     & $1.60765 \times 10^6$    & $3.30$         & $-10.26$               \\ 
        exp\cite{ph2solex:93oka}     & -          & -              & -                        & -              & $-11.43$               \\ 
        \hline\hline
    \end{tabular}
	
	\label{tab:shift_params}
\end{table*}

\begin{figure} [H]
	\centering
	\includegraphics[width=\linewidth]{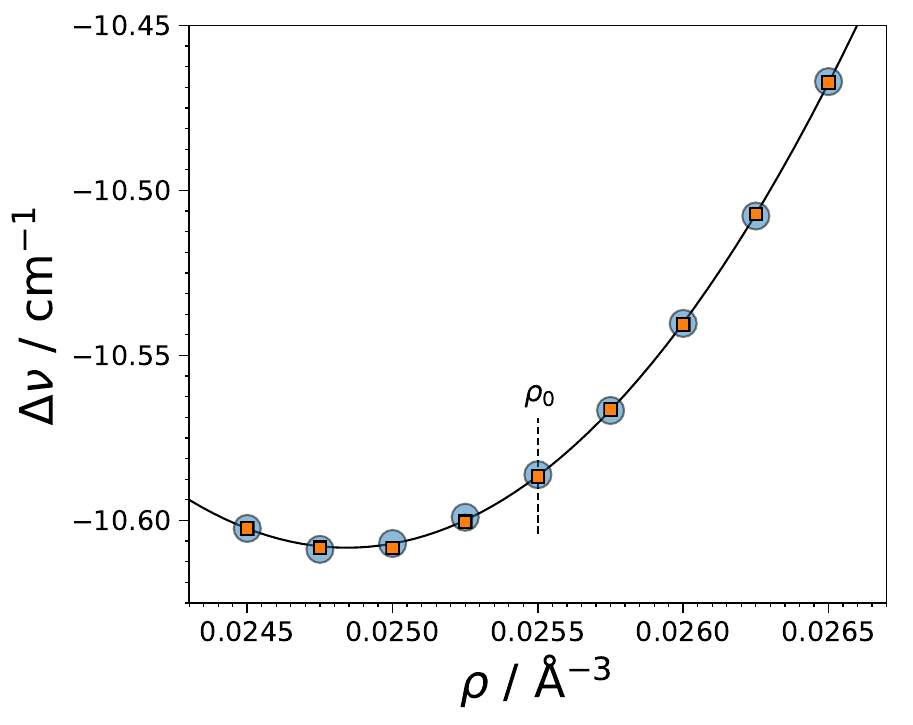}
	\caption{The vibrational matrix shift for the \hcp (blue circles) and \fcc (orange squares) crystals near the equilibrium density, calculated using Eq.~(\ref{eq:new_shift}) after performing the extrapolations $ \tau \rightarrow 0 $ and $ N \rightarrow \infty $. Both lattices provide essentially the same virational shift. The minimum value of $ \Delta \nu $ is found around $ \rho = 0.0248 \, \ANG^{-3} $, lower than the equilibrium density $ \rho_0 \approx 0.0255 \, \ANG^{-3} $ for the EOS indicated by the dashed line.}
	\label{fig:paper0_fig11}
\end{figure}

\begin{figure} [t]
	\centering
	\includegraphics[width=\linewidth]{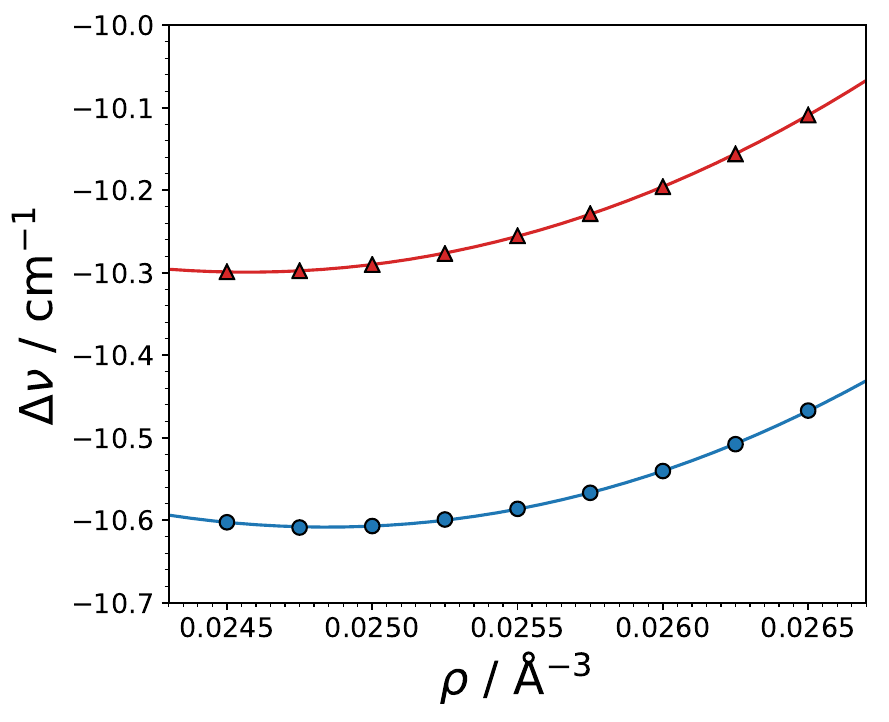}
	\caption{The vibrational matrix shift for an \hcp crystal with $ N = 180 $ \para molecules (red triangles) and with the extrapolation $ N \rightarrow \infty $ (blue circles). Performing the extrapolation lowers the value of $ \Delta \nu_0 $ from $ -10.26 $ cm$^{-1}$ to $ -10.59 $ cm$^{-1}$, and brings it closer to the experimental value of $ -11.43 $ cm$^{-1}$.}
	\label{fig:paper0_fig12}
\end{figure}

\begin{figure} [h]
	\centering
	\includegraphics[width=\linewidth]{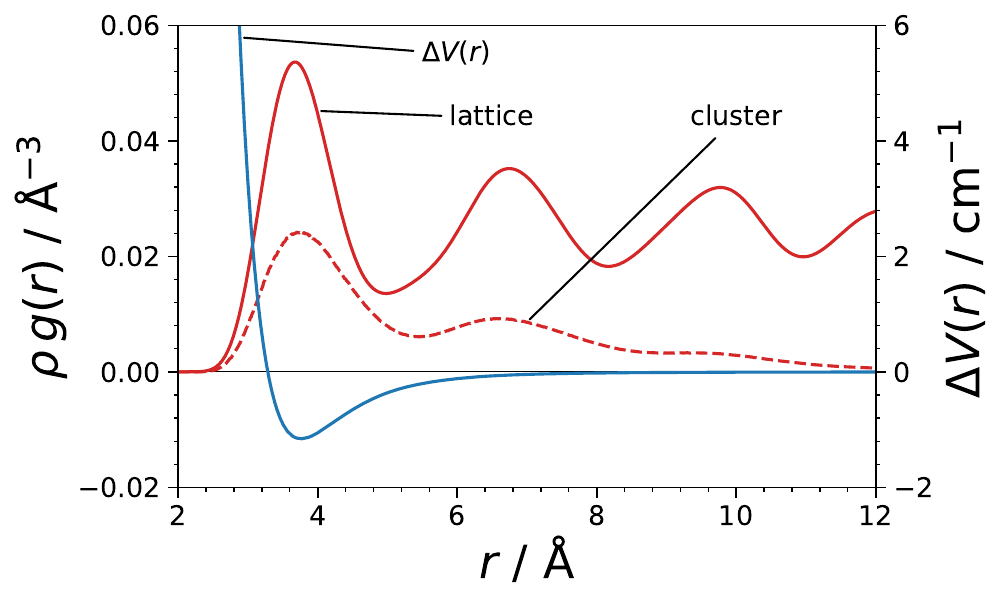}
	\caption{The normalized radial distribution $ \rho g(r) $ for an \hcp \para lattice (solid red line) and a cluster of $ N = 33 $ \para molecules (dashed red line), alongside the potential energy difference $ \Delta V(r) = V^1(r) - V^0(r) $ (solid blue line). The magnitude of the first shell of the \para lattice is roughly twice that of the \para cluster in the region where $ \Delta V(r) $ is the most negative. Note that while the radial distributions for the lattice and cluster were calculated the same way, the cluster does not have a well-defined density $ \rho $.}
	\label{fig:paper0_fig13}
\end{figure}

We show $ \Delta \nu_0 $, the vibrational band origin shift at the equilibrium density $ \rho_0 $, alongside the fit parameters in Tab.~\ref{tab:shift_params}. We see that the value of $ \Delta \nu_0 $ calculated in this report differs from the experimental value by only $ 8 $ \%. In Fig.~(\ref{fig:paper0_fig11}) we see that, once again, the vibrational matrix shifts for the \fcc and \hcp solids are essentially identical near the equilibrium density. We should thus expect that a measurement of $ \Delta \nu $ is an inadequate way to determine if a sample of \spara is in an \fcc or \hcp state.
The value of $ \Delta \nu $ in large-sized ($N > 10$) \para clusters was found to be between $ -3 $~cm$^{-1}$ and $ -4 $~cm$^{-1}$,\cite{ph2clu:04teje, ph2clu:14Nabi}~less than half of its value in \spara ($-11.43$~cm$^{-1}$).\cite{ph2solex:93oka}~To explain this discrepancy, in Fig.~(\ref{fig:paper0_fig13}) we can compare the radial distribution functions for a cluster of $ N = 33 $ \para molecules\cite{ph2clu:14Nabi} and an \hcp lattice of \para molecules with periodic boundary conditions applied.
The extrapolation of the system to infinite size is predicted by Eq.~(\ref{eq:new_shift}) to lower the vibrational frequency shift, because $ V^1(r) < V^0(r) $ at large distances. In Fig.~(\ref{fig:paper0_fig12}) we show $ \Delta \nu $~\vs~$ \rho $ for an \hcp lattice with $ N = 180 $ \para molecules and one having applied the extrapolation $ N \rightarrow \infty $. The latter is about $ 0.3 $~cm$^{-1}$ lower than the former, moving it closer to the experimental value of $ -11.43 $~cm$^{-1}$.\cite{ph2solex:93oka}

The $ g(r) $ curves for \spara for different values of $ N $ and $ \tau $ are indistinguishable on the scale of Fig.~(\ref{fig:paper0_fig13}), so the result for $ N = 1440 $ and $ P = 192 $ is chosen. Both radial distribution functions have been renormalized such that $ 4 \pi \int \dd r \, r^2 \rho g(r) = (N - 1) $ based on the coefficient of Eq.~(\ref{eq:new_shift}). According to Eq.~(\ref{eq:new_shift}), the value of $ \Delta \nu $ is determined by the integral of the product of $ \Delta V(r) = V^1(r) - V^0(r) $ and the normalized radial distribution $ g(r) $. We see in Fig.~(\ref{fig:paper0_fig13}) that the radial distribution function of the \para lattice is greater than that of the \para cluster for all distances. The differing values of $ \Delta \nu $ for the \para cluster and \para solid are simply a result of the different numbers of \para--\para pairs. This is consistent with what we see in Ref.~\citenum{ph2clu:14Nabi}, where the value of $ \Delta \nu $ for the cluster becomes more negative as the cluster size increases from $ 13 $ to $ 33 $.

\begin{figure} [h]
	\centering
	\includegraphics[width=\linewidth]{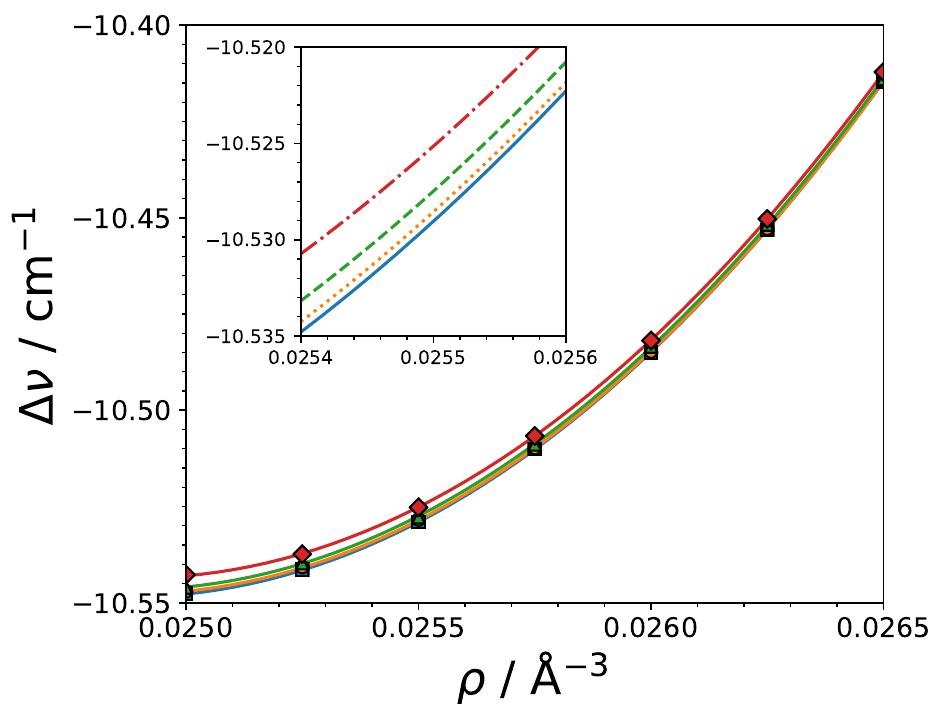}
	\caption{The vibrational matrix shift $ \Delta \nu $ for a \para \hcp crystal at temperatures of $ T = 2.10 $~K (blue squares, solid line in inset), $ T = 4.20 $~K (orange circles, dotted line in inset), $ T = 5.04 $~K (green triangles, dashed line in inset), and $ T = 6.30 $~K (red diamonds, dotted-dashed line in inset), with $ N = 448 $ and $ P $ adjusted such that in each simulation, $ \tau = \beta / P \approx 0.00248 $~K$^{-1}$.}
	\label{fig:paper0_fig14}
\end{figure}

In Fig.~(\ref{fig:paper0_fig14}) we plot $ \Delta \nu $ \vs $ \rho $ for temperatures $ T = \{ 2.10 \, K, 4.20 \, K, 5.04 \, K, 6.30 \, K \} $, with the number of time slices $ P $ adjusted such that $ \tau = \beta / P \approx 0.00248 $~K$^{-1}$ for each simulation. The curves are fit using Eq.~(\ref{eq:murnaghan}), and the fit parameters are provided in Tab.~\ref{tab:shift_temperature_params} next to their corresponding values of $ \Delta \nu_0 $. Similarly to the EOS isotherms of \para calculated with the FSH potential, the corresponding vibrational matrix shift isotherms also decrease with temperature, and the changes are nearly negligible below $ T = 4.2 $~K.

\begin{table} [ht]
	\centering
	\caption{Coefficients $a$, $b$, $c$, and obtained by fitting Eq.~(\ref{eq:murnaghan}) to the ($ \rho $, $ \Delta \nu $) pairs, for \hcp \para crystals at different temperatures. All simulations here are performed using $ \tau \approx 0.00254 $~K$^{-1}$ and $ N = 448 $ \para molecules. Each fit used a fixed exponent parameter $ \gamma = 3.30 $. Also included is $ \Delta \nu_0 $ (in cm$^{-1}$), the vibrational matrix shift at equilibrium density $ \rho_0 = 0.02546 \, \ANG^{-3}$.}

    \begin{tabular}{l@{\hskip 0.15in}l@{\hskip 0.15in}l@{\hskip 0.15in}l@{\hskip 0.15in}l}
        \hline\hline
        ~              & \thead{$a$} & \thead{$b$} & \thead{$c$}           & \thead{$\Delta \nu_0$} \\ \hline
        $ T = 2.10 $~K & $7.952$     & $-1067.07$  & $1.58275 \times 10^6$ & $-10.5312$               \\ 
        $ T = 4.20 $~K & $7.970$     & $-1067.87$  & $1.58318 \times 10^6$ & $-10.5307$               \\ 
        $ T = 5.04 $~K & $7.908$     & $-1064.28$  & $1.57800 \times 10^6$ & $-10.5296$               \\ 
        $ T = 6.30 $~K & $8.035$     & $-1070.83$  & $1.58573 \times 10^6$ & $-10.5272$               \\ 
        \hline\hline
    \end{tabular}
	
	\label{tab:shift_temperature_params}
\end{table}

\section{Conclusion\label{sec:conclusion} }

We have presented extensive PIMC studies of the EOS curves of both \fcc and \hcp structured \spara based on a first-principles AHR effective pair potential for the \para dimer. To avoid the systematic errors associated with the finite size of the simulation cell and the finite value of imaginary time step, the results are extrapolated in the limits $ N \rightarrow \infty $ and $ \tau \rightarrow 0 $. The calculated energy per particle as a function of density is then fit to a Murnaghan-type curve. The resulting EOS curves are compared with previously reported results, and the discrepancies are explained by differences in the two-body \para--\para PESs and simulation methodologies. In addition, the equilibrium density obtained and the corresponding equilibrium energy per particle agree very well with experimental observations. The \hcp structured \spara crystal is shown to be more stable than the \fcc structured one. The equilibrium density of the EOS curves was found to not vary with temperature. We also calculated the pressure as a function of density and the compressibility as a function of pressure, based on the obtained EOS curves, and provide a reasonable analysis of the results.

Furthermore, this work reports the first ever \apriori prediction of a matrix vibrational band origin shift. This shift is predicted to be a marked function of density, and the value at the equilibrium density is quite close to the experimental result. The highly accurate EOS obtained should provide a reliable platform for further studies on the structural and spectroscopic properties of both pure and doped \spara.

\section*{Acknowledgements}
The authors acknowledge the Natural Sciences and Engineering Research Council (NSERC) of Canada (RGPIN-2016-04403), the Ontario Ministry of Research and Innovation (MRI), the Canada Research Chair program (950-231024), and the Canada Foundation for Innovation (CFI) (project No. 35232).


\bibliography{./pureh2}

\begin{thebibliography}{10}

\bibitem{ph2solex:66nosa}
L.~H. Nosanow, Phys.\ Rev. {\bf 146},  120  (1966).

\bibitem{ph2solex:80silv}
I.~F. Silvera, Rev.\ Mod.\ Phys. {\bf 52},  393  (1980).

\bibitem{ph2solex:60gush}
H.~P. Gush, W.~F.~J. Hare, E.~J. Allin, and H.~L. Welsh, Can.\ J.\ Phys. {\bf
  38},  176  (1960).

\bibitem{ph2solex:66barr}
C.~S. Barrett, L. Meyer, and J. Wasserman, J.\ Chem.\ Phys. {\bf 45},  834
  (1966).

\bibitem{ph2solex:67bost}
O. Bostanjoglo and R. Kleinschmidt, J.\ Chem.\ Phys. {\bf 46},  2004  (1967).

\bibitem{ph2solex:70schne}
O. Schnepp, Phys.\ Rev.\ A {\bf 2},  2574  (1970).

\bibitem{ph2solth:91Zoppi}
M. Zoppi and M. Neumann, Phys.\ Rev.\ B {\bf 43},  10242  (1991).

\bibitem{ph2solex:92momo}
T. Momose, D.~P. Weliky, and T. Oka, J.\ Mol.\ Spectrosc. {\bf 153},  760
  (1992).

\bibitem{ph2solex:93oka}
T. Oka, Ann.\ Rev.\ Phys.\ Chem. {\bf 44},  299  (1993).

\bibitem{ph2solth:02Zoppi}
M. Zoppi, M. Neumann, and M. Celli, Phys.\ Rev.\ B {\bf 65},  092204  (2002).

\bibitem{ph2solth:06Oper}
F. Operetto and F. Pederiva, Phys.\ Rev.\ B {\bf 73},  184124  (2006).

\bibitem{ph2solth:06Tom}
T. Lindenau {\it et~al.}, Int.\ J.\ Mod.\ Phys.\ B {\bf 20},  5035  (2006).

\bibitem{suph2:00greb}
S. Grebenev, B.~G. Sartakov, J.~P. Toennies, and A.~F. Vilesov, Science {\bf
  289},  1532  (2000).

\bibitem{suph2:01agreb}
S. Grebenev {\it et~al.}, Faraday Disc. {\bf 118},  19  (2001).

\bibitem{suph2:01bgreb}
S. Grebenev, B.~G. Sartakov, J.~P. Toennies, and A.~F. Vilesov, J.\ Chem.\
  Phys. {\bf 114},  617  (2001).

\bibitem{suph2:02greb}
S. Grebenev, B.~G. Sartakov, J.~P. Toennies, and A.~F. Vilesov, Phys.\ Rev.\
  Lett. {\bf 89},  225301  (2002).

\bibitem{suph2:03greb}
S. Grebenev, B.~G. Sartakov, J.~P. Toennies, and A.~F. Vilesov, J.\ Chem.\
  Phys. {\bf 118},  8656  (2003).

\bibitem{suph2:05peas}
F. Paesani, R.~E. Zillich, Y. Kwon, and K.~B. Whaley, J.\ Chem.\ Phys. {\bf
  122},  181106  (2005).

\bibitem{suph2:10greb}
S. Grebenev, B.~G. Sartakov, J.~P. Toennies, and A.~F. Vilesov, J.\ Chem.\
  Phys. {\bf 132},  064501  (2010).

\bibitem{suph2:10hli}
H. Li, R.~J. Le~Roy, P.-N. Roy, and A.~R.~W. McKellar, Phys.\ Rev.\ Lett. {\bf
  105},  133401  (2010).

\bibitem{suph2:13atoby}
T. Zeng, H. Li, and P.-N. Roy, J.\ Phys.\ Chem.\ Lett. {\bf 4},  18  (2013).

\bibitem{suph2:13btoby}
T. Zeng, G. Guillon, J.~T. Cantin, and P.-N. Roy, J.\ Phys.\ Chem.\ Lett. {\bf
  4},  2391  (2013).

\bibitem{ph2solth:12Osy}
O.~N. Osychenko, R. Rota, and J. Boronat, Phys.\ Rev.\ B {\bf 85},  224513
  (2012).

\bibitem{doh2solex:93weli}
D.~P. Weliky {\it et~al.}, J.\ Chem.\ Phys. {\bf 105},  4461  (1996).

\bibitem{doh2solex:98zhang}
Y. Zhang {\it et~al.}, Phys.\ Rev.\ B {\bf 58},  218  (1998).

\bibitem{doh2solex:04momo}
T. Momose, H. Honshina, M. Fushitani, and H. Katsuki, Vib.\ Spectrosc. {\bf
  34},  95  (2004).

\bibitem{doh2solex:13faja}
M.~E. Fajardo, J.\ Phys.\ Chem.\ A {\bf 117},  13504  (2013).

\bibitem{ph2clu:04teje}
G. Tejeda {\it et~al.}, Phys.\ Rev.\ Lett. {\bf 92},  223401  (2004).

\bibitem{ph2clu:14Nabi}
N. Faruk {\it et~al.}, J.\ Chem.\ Phys. {\bf 141},  014310  (2014).

\bibitem{ph2solex:63good}
R.~D. Goodwin and H.~M. Roder, Cryogenics {\bf 3},  12  (1963).

\bibitem{ph2solex:75dura}
S.~C. Durana and J.~P. McTague, J.\ Low Temp.\ Phys. {\bf 21},  21  (1975).

\bibitem{ph2solth:78silv}
I.~F. Silvera and V.~V. Goldman, J.\ Chem.\ Phys. {\bf 69},  4209  (1978).

\bibitem{ph2solex:79drie}
A. Driessen, J.~A. de~Waal, and I.~F. Silvera, J.\ Low Temp.\ Phys. {\bf 34},
  255  (1979).

\bibitem{ph2solex:68schu}
A.~F. Schuch, R.~L. Mills, and D.~A. Depatie, Phys.\ Rev. {\bf 165},  1032
  (1968).

\bibitem{ph2solth:96chen}
E. Cheng and K.~B. Whaley, J.\ Chem.\ Phys. {\bf 104},  3155  (1996).

\bibitem{h2pes:78sg}
I.~F. Silvera and V.~V. Goldman, J.\ Chem.\ Phys. {\bf 69},  4209  (1978).

\bibitem{h2pes:83buck}
U. Buck {\it et~al.}, J.\ Chem.\ Phys. {\bf 78},  4439  (1978).

\bibitem{h2pes:08hinde}
R.~J. Hinde, J.\ Chem.\ Phys. {\bf 128},  154308  (2008).

\bibitem{ph2clu:15schm}
M. Schmidt {\it et~al.}, J.\ Phys.\ Chem.\ A {\bf 119},  12551  (2015).

\bibitem{pi:14tobya}
T. Zeng and P.-N. Roy, Rep.\ Prog.\ Phys. {\bf 77},  046601  (2014).

\bibitem{rovib:59kran}
J. can Kranendonk, Physica {\bf 25},  1080  (1959).

\bibitem{rovib:68kran}
J. van Kranendonk and G. Karl, Rev.\ Mod.\ Phys. {\bf 40},  531  (1968).

\bibitem{adiab:09krz}
K. Pachucki and J. Komasa, J.\ Chem.\ Phys. {\bf 130},  164113  (2009).

\bibitem{pi:16tobyb}
T. Zeng {\it et~al.}, Comp.\ Phys.\ Comm. {\bf 204},  170  (2016).

\bibitem{pi:95cepe}
D.~M. Ceperley, Rev.\ Mod.\ Phys. {\bf 67},  279  (1995).

\bibitem{dhe:09hlia}
H. Li, N. Blinov, P.-N. Roy, and R.~J. Le~Roy, J.\ Chem.\ Phys. {\bf 130},
  144305  (2009).

\bibitem{dhe:12wang}
L. Wang, D. Xie, R.~J. Le~Roy, and P.-N. Roy, J.\ Chem.\ Phys. {\bf 137},
  104311  (2012).

\bibitem{pimc:17yan}
Y. Yan and D. Blume, J.\ Phys.\ B: At.\ Mol.\ Opt.\ Phys. {\bf 50},  223001
  (2017).

\bibitem{dhe:11wang}
L. Wang {\it et~al.}, J.\ Mol.\ Spectrosc. {\bf 267},  136  (2011).

\bibitem{pbc:04yeh}
I.-C. Yeh and G. Hummer, J.\ Phys.\ Chem.\ B {\bf 108},  15873  (2004).

\bibitem{pbc:09naka}
H. Nakano and A. Terai, J.\ Phys.\ Soc.\ Jpn. {\bf 78},  014003  (2009).

\bibitem{h2pes:10hli}
H. Li, P.-N. Roy, and R.~J. Le~Roy, J.\ Chem.\ Phys. {\bf 133},  104305
  (2010).

\bibitem{tail:87allen}
M.~P. Allen and D.~J. Tildesley, {\em Computer Simulation of Liquids} (Oxford
  University Press, New York, 1987).

\bibitem{fiteq:44murn}
F.~D. Murnaghan, Proc.\ Nat.\ Acad.\ Sci.\ (US) {\bf 30},  244  (1944).

\bibitem{h2h2h2pes:08hinde}
R.~J. Hinde, Chem.\ Phys.\ Lett. {\bf 460},  141  (2008).

\bibitem{compr:70udov}
B.~G. Udovidchenko and V.~G. Manzhelii, J.\ Low Temp.\ Phys. {\bf 3},  429
  (1970).

\end{thebibliography}
\bibliographystyle{prsty}

\newpage


\newpage


\end{document}